# Evaluation of Microscopic Origins in Nonlinear Optical Crystals: Based on Rigorous Atomic Space Tessellating

Yang Chi*

**ABSTRACT:** Clarifying the contribution of various atoms and structural units formed in nonlinear optical (NLO) crystal materials to macroscopic optical response is crucial for NLO crystal design. In this work, the rigorous atomic space tessellating (AST) method without any empirical parameters is introduced to evaluate the correlated contributions in wavelength dependent optical responses. The NLO crystal $CsSbF_2SO_4$ with significant asymmetric electron distribution of $Sb^{3+}$ was selected as the case study. This study indicates that there are significant differences in atomic and dipole moment contributions between static and dynamic second harmonic generation (SHG) susceptibility. Using only the information obtained from NLO-active unit analysis of static SHG susceptibilities to explain the NLO phenomenon in actual photoexcitation can even lead to qualitative errors. Among all the transitions in SHG effects, the on-site transition of Sb has the largest component, while the O−Sb off-site transition with multiple transition channels contributes the most to the macroscopic SHG susceptibilities. The relevant methods have also been extended to linear optics and used to study first-order susceptibilities anisotropy, which is important for achieving phase matching. $Sb^{3+}$ contributes significantly to the linear optical anisotropy with the same sign as $SO_4^{2-}$.

## INTRODUCTION

Nonlinear optical (NLO) crystals play a fundamental role in the widespread application of laser technology in modern science and technology. Developing NLO crystals with excellent linear optical (LO) and NLO properties, and a wider transmission range, especially in deep ultraviolet, mid- to far-infrared, and even terahertz spectral regions, remains an attractive topic today, more than six decades after the invention of lasers. Many researchers working in this field are still striving to find new NLO crystals.

For a crystal to exhibit second harmonic generation (SHG) response, it must possess broken time- and/or space-inversion symmetry[1]. Therefore, constructing crystalline materials with symmetry breaking by combining units that exhibit broken symmetry is a simple and effective strategy[2]. It is evident that by evaluating the microscopic origins of LO and NLO effects in crystals from different atoms/groups, we can gain a deeper understanding of the structure-property relationships of materials, which can provide valuable information for the exploration of new NLO crystals.

In order to comprehensively understand the mechanisms of NLO response, researchers have proposed various methods to evaluate the contributions of different atoms and groups in NLO crystals. In mid-1970s, the anionic group model was introduced by Chen[3], which calculates the second-order susceptibility using the localized molecular orbital method under the first-order approximation. This theory yielded results that were consistent with experiments in crystals such as $\beta$-BaB$_2$O$_4$.

Subsequently, several theoretical schemes were successively proposed, each with its own unique characteristics. For instance, the band-resolved analysis method allows for the determination of contributions from different bands[4]. The real-space atom-cutting method[5] with semiempirical parameters enables the evaluation of SHG contributions from individual atoms or groups through the partitioning of wavefunctions, and has been applied in numerous studies[6]. However, this method is unable to handle atoms that form strong covalent bonds (such as ions in borate groups) and exhibit non-spherical electron distributions (for instance, cations containing lone pair electrons)[7]. Furthermore, the inherent limitations of this method include the reuse of boundary atoms or the loss of transition matrix elements between their orbitals[8]. Additionally, the method cannot effectively evaluate the unlocal NLO effects[5]. Therefore, the macroscopic SHG susceptibility cannot be strictly divided into specific atoms or groups from this approach. The SHG-weighted electron density analysis developed afterwards can visually identify the orbital distribution that contributes significantly to the SHG susceptibility, and has also gained widespread application[9].

Recently, several methods based on atomic orbital projection have undergone rapid development. The atomic response theory has been developed to evaluate the contributions of individual atom to SHG response based on the band projected coefficients of atomic orbitals. However, the differences in contributions from Bloch wavevectors are treated by averaging[10].

In a recent study, Bloch states were projected onto the atomic-like Wannier orbitals to obtain projection coefficients, and local Wannier was used to analyze local orbit contributions to the SHG susceptibilities[11]. Nevertheless, the SHG response inherently involves complex photoexcitation and radiation processes that encompass both off-site and on-site transitions, transcending the confines of a single atom. More recently, there have been studies using atomic orbital projection SHG technique to analyze key dipole moments in the SHG process, and also discussed the variation of atomic contribution with incident wavelength[12].

Microscopic contribution analysis methods involving atomic orbital projection also confront the issue of inability to strictly partition SHG contributions, as the projected functions are not orthonormal[13]. In addition, schemes involving projected density of states (PDOS) can lead to incomplete allocation, as truncation of Wigner-Seitz radius results in interstitial density. These factors contribute to the inability of methods involving atomic orbital projection to ensure that the global SHG susceptibility is strictly allocated to all atoms or groups.

Given the current experimental and theoretical landscape, a novel approach to evaluate the microscopic contributions of SHG should encompass all the advantages of existing methods while avoiding their shortcomings.

In this study, by comprehensively considering the contributions and weights of Bloch states in both momentum and energy spaces, a method based on rigorous atomic space tessellating (AST) has been developed to assess the microscopic origins of wavelength-dependent SHG effects in NLO crystals. This has led to the development of various related analysis techniques that encompass all the advantages of the aforementioned methods while avoiding their drawbacks.

The fluoride sulfate $CsSbF_2SO_4$ (CSFS) with active lone pair cations which is phase-matchable and revealed a strong SHG response in metal sulfate is used as a case study[14]. The transitions related to Sb have the greatest impact on the SHG effect, especially the on-site transitions of Sb and the off-site transitions from O to Sb. Sb also contributes significantly to the refractive index anisotropy of CSFS.

**RESULTS AND DISCUSSION**

The CSFS structure crystalizing in the polar *Pna*2$_1$ space group (achiral-polar point group *C2v*), has unique Cs1, Sb1 and S1 atom sites, as well as two F atom sites (F1 and F2) and four O atom sites (O1–O4) in asymmetric unit. For the SHG in a system with filled bands, expressions with scissor correction derived within the length-gauge formalism used to calculate the frequency-dependent SHG susceptibility $\chi^{(2)}_{ijk}(-2\omega; \omega, \omega)$ were originally derived by Aversa and Sipe[15] and later modified by F. Nastos *et al*[16]. This method typically yields results that are very consistent with experimental values[17]. On the basis of the point group *mm*2, five independent SHG susceptibility tensor elements $\chi^{xxz}$, $\chi^{yyz}$, $\chi^{zxx}$, $\chi^{zyy}$, and $\chi^{zzz}$ were calculated (**Figure S1**). The calculated ⟨$d_{\text{eff}}$⟩ values for the powder sample are 1.18 pm/V at 1064 nm, which are in good agreement with experimental results (~3 × KDP).

However, solely relying on the macroscopic SHG susceptibility cannot distinguish the roles played by different atomic and structural units in the NLO process, nor can it reveal the underlying microscopic structure-activity mechanisms. Bloch electrons are not localized around a single atom but are distributed throughout the entire crystal and shared by all atoms. Therefore, it is not straightforward to define or obtain the microscopic distribution of SHG related to specific atoms or groups using the above expressions. The rigorous AST method provides a potential avenue for comprehensively and multi-dimensionally analyzing microscopic contributions.

According to Eqs. (S1–S3)[16], $\chi^{(2)}_{ijk}$ can be decomposed as the sum of contributions from different Bloch states. Firstly, the Bloch states at different *k*-points are sorted from low to high in energy and denoted as $E_i(\mathbf{k}_j)$. The influence on the SHG susceptibility dominated by a specific state with band index *j* and wavevector $\mathbf{k}_j$ is defined as,

$$\chi^{abc,\mathbf{k}_j}_{E_i(\mathbf{k}_j)} = \frac{1}{2}\left[\frac{1}{1+e^{\frac{E_i(\mathbf{k}_j)-E_F}{kT}}}\left(\chi^{abc,\mathbf{k}_j}_{n,m,l\geq i} - \chi^{abc,\mathbf{k}_j}_{n,m,l>i}\right) + \frac{1}{1+e^{\frac{E_F-E_i(\mathbf{k}_j)}{kT}}}\left(\chi^{abc,\mathbf{k}_j}_{n,m,l\leq i} - \chi^{abc,\mathbf{k}_j}_{n,m,l<i}\right)\right]. \quad (1)$$

As quantities existing in the complex plane, when the zero-frequency limit is taken, the SHG susceptibility degenerates from the complex to the real field, meaning that its imaginary part becomes zero. At this point, the magnitude and sign of each Bloch state's contribution to the macroscopic susceptibility become straightforward. As a result, nearly all research regarding the microscopic contributions to the SHG stems from the analysis of static susceptibility[4–11].

To determine the contribution of each Bloch state under non-zero frequency optical excitation, it is necessary to consider the differences in the contributions of different $k$-points to the total SHG susceptibility (**Figure 1a**), as well as the sum rules in the SHG calculation expression. The contribution of each Bloch state to the macroscopic SHG susceptibility must undergo two projections: first, projection onto the total susceptibility for each $k$-point, and then, projection onto the macroscopic SHG susceptibility (**Figure 1**). And naturally satisfies the following equation,

$$\chi^{abc} = \sum_{\mathbf{k}} \sum_{E} \chi_{E_i(\mathbf{k}_j)}^{abc,\mathbf{k}_j} = \sum_{\mathbf{k}} \sum_{E} \text{proj}_{\chi^{abc}} \left[ \text{proj}_{\chi^{abc,\mathbf{k}_j}} \left( \chi_{E_i(\mathbf{k}_j)}^{abc,\mathbf{k}_j} \right) \right]. \quad (2)$$

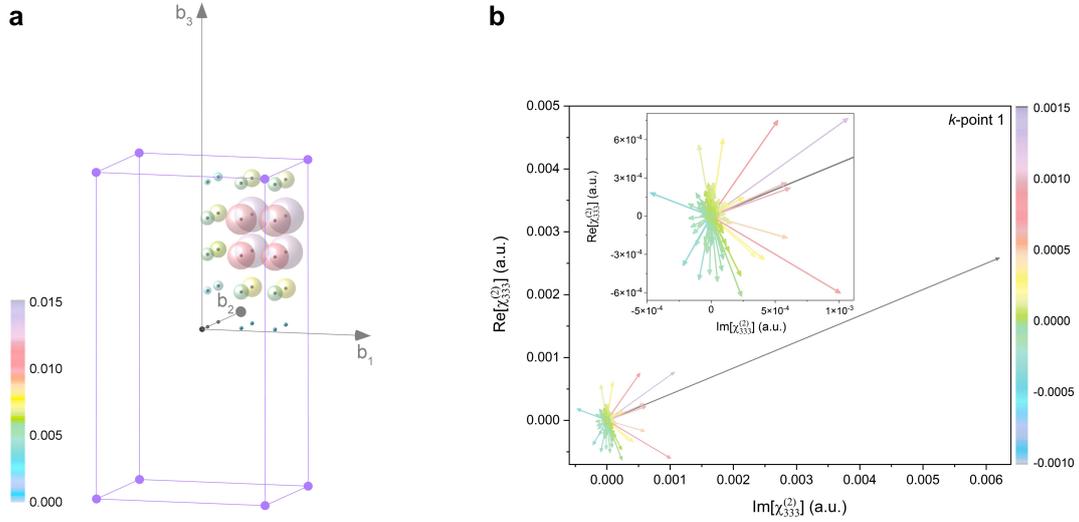

**Figure 1.** (**a**) In the reciprocal space of CSFS, the proportional relationship between the SHG components of the 30 irreducible $k$-points is represented by the radius of the spheres, while the magnitudes of their projections onto the macroscopic SHG susceptibility $\chi^{zzz}$ are indicated by the color of the spheres (units in a.u.). (**b**) The relationship between the SHG contributions dominated by different Bloch states and the magnitudes of their projections onto the component $\chi^{zzz,\mathbf{k}_1}$ (represented by a color scale, units in a.u.). These data are all for the incident wavelength of 1064 nm.

Atomic space is the local space attributed to specific atoms in the whole three-dimension space. Different atomic space partitions correspond to different weight functions $w_A$. The methods for partitioning direct space into different atomic spaces can be divided into two categories: discrete and fuzzy partition methods. A comparison between discrete and fuzzy partition methods shows that the difference between the two methods lies in the choice of the weight factor. The fuzzy method defines a sort of fuzzy atom whereas the discrete partition method determines sharp boundaries between the atoms.

To ensure the atomic space partitioning without empirical parameters, Voronoi[18] and atoms in molecules[19] (AIM) partitions (also known as QTAIM or Bader's partition) in discrete partition methods as well as Hirshfeld[20] and Hirshfeld-I[21] partitions in fuzzy partition methods are selected in this study. Additional details are in the Supporting Information.

To obtain the microscopic SHG susceptibilities for different atoms, the SHG contributions of each Bloch state in the momentum and energy spaces are projected onto the respective atomic spaces of the corresponding Bloch orbitals. This enables us to determine the contribution of any orbital in the momentum and energy spaces to the SHG susceptibility at any point within any atomic space. Thus, the atomic space SHG distribution function is defined as follows,

$$^{\text{proj}}\chi^{abc,\mathbf{k}_j}_{E_i(\mathbf{k}_j),w_A}(\mathbf{r}) = \text{proj}_{\chi^{abc}}\left\{\text{proj}_{\chi^{abc,\mathbf{k}_j}}\left[\chi^{abc,\mathbf{k}_j}_{E_i(\mathbf{k}_j)}\varphi^2_{E_i(\mathbf{k}_j)}(\mathbf{r})w_A(\mathbf{r})\right]\right\}. \quad (3)$$

All atomic spaces can be considered as generalized partitioning schemes of three-dimensional tessellation,

$$\chi^{abc} = \sum_{\mathbf{k}}\sum_{E}\sum_{w_A}\int {}^{\text{proj}}\chi^{abc,\mathbf{k}_j}_{E_i(\mathbf{k}_j),w_A}(\mathbf{r})\,d\mathbf{r}. \quad (4)$$

That is to say, all contributions can be strictly divided (i.e., the sum of all integral value of atomic space SHG distribution function is strictly equal to macroscopic SHG) without any empirical parameters, which is a significant advancement. It is worth mentioning that this method can be further enriched and expanded by changing the atomic space weight functions $w_A$.

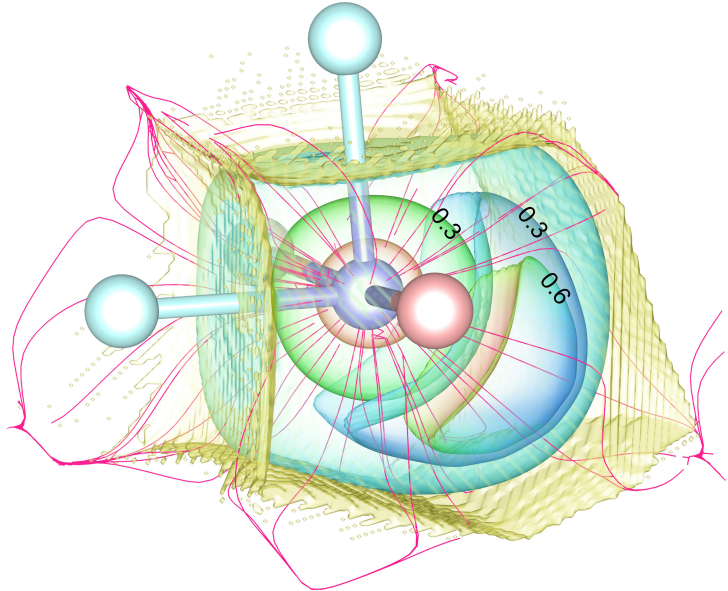

**Figure 2.** The AIM atomic space of Sb1 (enclosed by a yellow surface), sum of all the SHG distribution function in this atomic space $^{\text{proj}}\chi^{zzz}_{w_{\text{AIM-Sb1}}}(\mathbf{r})$ at an incident wavelength of 1064 nm (green isosurface, isosurface level = 0.0004 Å$^{-3}$, units in a.u.), and the ELF isosurface in this atomic space (isosurface of surface coloring, isosurface level = 0.3 and 0.6). The rose red line is used to trace gradient paths originating at a Sb1 atom.

The AIM and Voronoi atomic spaces belonging to discrete partition approach can be well attributed to the contributions of related atoms, even for ions such as Sb$^{3+}$ that contain lone pair electrons and are deviates from roundness. By presenting the electron localization function[22] (ELF) as well as SHG distribution function, atomic bases, and electron density gradient lines of Sb AIM atomic spaces together, it can be clearly seen that the Sb AIM atomic space and sum of all the SHG distribution function in it $^{proj}\chi^{zzz}_{w_{AIM-Sb1}}(\mathbf{r})$ are well consistent with the local electron distribution displayed by ELF (**Figure 2**).

If a simplistic spherical approach is adopted (e.g., using the bond critical points as the reference radius of the sphere, but adjustments are needed to avoid spatial overlap, **Figure 3a** and **S2**), not only will the results fail to be normalized, but the contribution of Sb's lone pair electrons will also be overlooked (**Figure 3a**). On the other hand, if the radius of Sb is increased (e.g., to 1.87 Å), significant overlap will occur with the coordinating atoms O and F (**Figure 3a**). Moreover, for atomic groups with strong covalent interactions, the AIM atomic space can also be reasonably partitioned (**Figure 3b** and **c**).

Comparing the AIM and Voronoi atomic spaces, it can be observed that their spatial partitioning results are very similar, with the Voronoi cell-like atomic space appearing as a convex polyhedral approximation of the AIM atomic space (**Figure 3**). This partitioning method can therefore also effectively separate the contributions of various types of atoms. Interestingly, the AIM atomic spaces of metal cations ($Sb^{3+}$ and $Cs^+$) are generally slightly larger than their Voronoi atomic spaces, while the opposite is true for non-metal anions ($O^{2-}$ and $F^-$) (**Figure 3**).

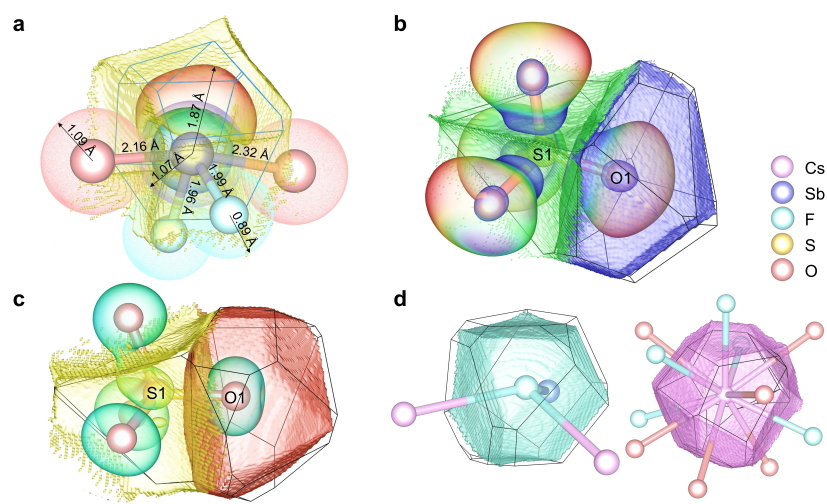

**Figure 3.** (**a**) The spatial relationship between the AIM and Voronoi atomic spaces of Sb1, as well as the ELF isosurface located inside them (isosurface level = 0.3). The radius of the spherical point cloud is set based on the chemical environment and bond critical points around Sb. (**b**) The AIM and Voronoi atomic spaces of S1 and O1 as well as ELF isosurface located inside S/O atoms spaces (isosurface level = 0.3). (**c**) Sum of all the SHG distribution function in these atomic spaces, i.e., $^{proj}\chi^{zzz}_{w_{AIM-SO_4}}(\mathbf{r})$, at an incident wavelength of 1064 nm (isosurface level = 0.0007 Å$^{-3}$, units in a.u.), together with the AIM and Voronoi atomic spaces of S1 and O1. (**d**) The AIM and Voronoi atomic spaces of F1 and Cs1.

For fuzzy partition approach, all atomic spaces are distributed throughout the entire space, and the weight function continuously varies within the range of [0, 1] as the spatial coordinates change. Comparing the different isosurfaces of the weight function for the fuzzy atom space (Hirshfeld and Hirshfeld-I) with the discrete atom space (AIM and Voronoi), it can be observed that when the isosurface level of the weight function is 0.8, the fuzzy atom space can be largely accommodated within the AIM and Voronoi spaces. As the isosurface level increases, the isosurface shrinks rapidly (**Figure 4**). Another notable point is that the isosurface shape of fuzzy atom space weight function exhibits a high degree of consistency with the shape of the partitioned atom space, including the pyramid-like shape of S atomic space (**Figure 4**).

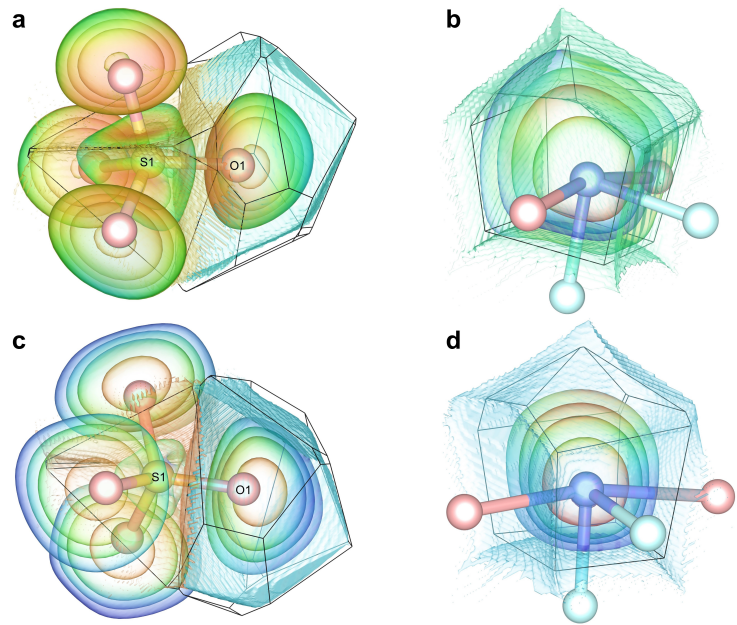

**Figure 4.** The Hirshfeld (**a** and **b**) and Hirshfeld-I (**c** and **d**) fuzzy atomic space weight function isosurfaces of SO$_4$ group (**a** and **c**) and Sb (**b** and **d**), as well as the AIM and Voronoi atomic spaces of S1, O1, and Sb1. All the subfigures contain four isosurfaces, with contour levels of 0.8, 0.9, 0.95, and 0.99, respectively.

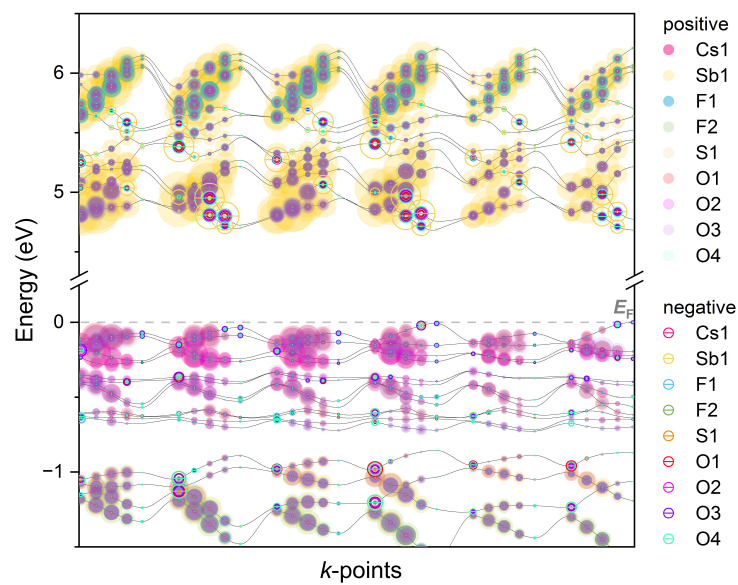

**Figure 5.** SHG-fatband plot of tensor element $\chi^{zzz}$ using AIM partition at an incident wavelength of 1064 nm. Positive and negative components are represented by different symbols.

By projecting the integral value of the atomic space SHG distribution function onto the Bloch states, the SHG response of atomic contribution distribution with energy and *k*-space resolution (later referred to as SHG-fatband) can be obtained. This allows for the subsequent SHG-fatband analysis of Bloch states of interest. **Figure 5** shows the SHG-fatband plot of $\chi^{zzz}$ obtained using the AIM atomic space partitioning approach for CSFS at an incident wavelength of 1064 nm. The SHG contributions from the empty band energy region near the Fermi level mainly originate from Sb and O atoms, while the valence band contributions are almost entirely from O atoms. Most regions generate positive components, with only a few Bloch states, including those near the valance band maximum (VBM) and conduction band minimum (CBM), producing negative components.

Taking into account that all SHG processes inevitably involve both valence and conduction bands, we can therefore obtain the transition components from all atomic pairs in the crystal through atomic space partitioning,

$$^{\text{proj}}\chi^{abc}_{A_1 \to A_2} = \sum_{\mathbf{k}} \sum_{E} \int {}^{\text{proj}}\chi^{abc,\mathbf{k}_j}_{w_{A_1,E_m^v}, w_{A_2,E_n^c}}(\mathbf{r}) \, d\mathbf{r}. \qquad (5)$$

All transition components obtained through this method also strictly satisfy,

$$\chi^{abc} = \sum {}^{\text{proj}}\chi^{abc}_{A_1 \to A_2}. \qquad (6)$$

In CSFS, the asymmetric unit contains 9 atoms, therefore, there are $9^2$ transitions between the relevant atoms, which can be used to plot a dipole moment component matrix diagram of tensor element $\chi^{zzz}$ at an incident wavelength of 1064 nm (**Figure 6**). It can be observed that the consistency of transition components obtained from these four atomic space partitioning methods are well-aligned, especially for AIM, Voronoi, and Hirshfeld-I methods, where the overall trends of the transition components are almost identical. Only the partitioning results of the Hirshfeld method slightly differ from the first three in some details. For example, in the first three methods, the contribution of the S1–Sb1 transition is less than that of F2–Sb1, while in the Hirshfeld method, it is slightly larger. Additionally, the transition components related to Cs1 are very small in the first three partitioning methods, but the Cs1–Sb1 and Sb1–Cs1 transition components are significantly higher in the Hirshfeld method. This discrepancy is likely due to the fact that the Hirshfeld method defines atomic spaces using free atomic states that do not respond to the surrounding chemical environment, and therefore may not be physically ideal. As a possible improvement, the Hirshfeld-I method introduces an iterative scheme based on the Hirshfeld method to refine the atomic spaces. The finally converged Hirshfeld-I atomic spaces appear to be more physically meaningful and eliminate the dependence on the initial choice of free states. When comparing the scaled atomic spaces with the original Hirshfeld atomic spaces, it is evident that the atomic spaces of cations collapse towards the positions of their nuclei, while the atomic spaces of anions expand significantly (**Figure 4**). Such a result naturally leads to changes in the distribution of SHG contributions.

In simple terms, the on-site transition of Sb contributes the most to the SHG effect, followed by the $O_i$–Sb1 off-site transitions. However, since there are more transition channels for O−Sb1, the $O_i$–Sb1 transitions contribute more to the macroscopic SHG susceptibility. It is worthy of mentioning that the contributions of transitions involving F atoms to the SHG susceptibility are smaller than those involving O atoms.

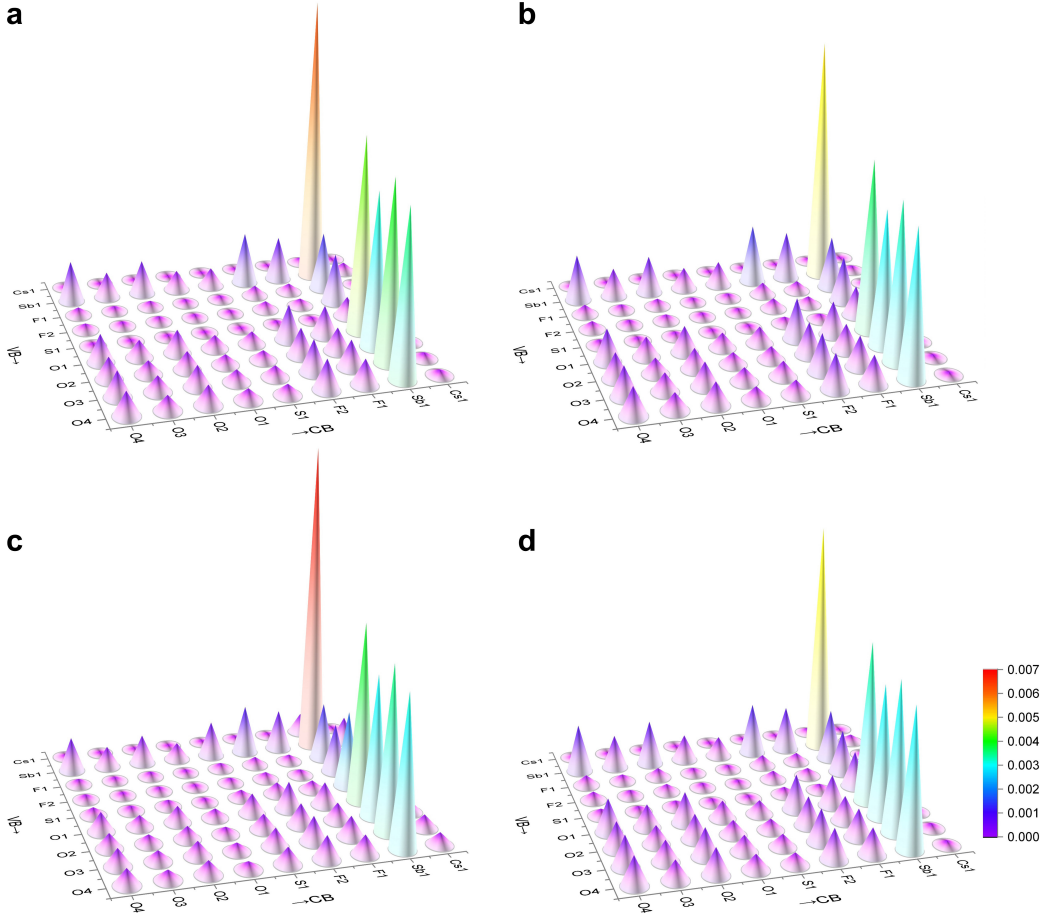

**Figure 6.** Atomic space dipole moment component matrix diagram for $\chi^{zzz}$ at an incident wavelength of 1064 nm, obtained through four atomic space partitioning methods, i.e., AIM (**a**), Vonoroi (**b**), Hirshfeld (**c**), and Hirshfeld-I (**d**), respectively. All units are atomic units.

After considering all the transition channels and lattice multiplicities within the unit cell, the SHG components from atoms at different lattice positions can be derived from their transition contributions (**Table 1**). The components from all atomic spaces are still satisfied,

$$\chi^{abc} = \sum{}^{\text{proj}} \chi_A^{abc}. \qquad (7)$$

$Sb^{3+}$ with active lone pair electrons indeed contributes significantly to SHG tensor element $\chi^{zzz}$ (**Table 1**), which explains why near-ultraviolet NLO material CSFS has a higher SHG response in metal sulfate.

It is worth noting that both experiments[23] and theories have shown that under different incident wavelengths, not only does the macroscopic SHG susceptibility change, but the relative change in the induced dipole moment also leads to changes in the contribution ratio of different atoms to the macroscopic SHG susceptibility. In the energy window of 0–2 eV, as the energy increases, the macroscopic SHG susceptibility of CSFS also increases. The energy-dependent atomic space contribution distribution (**Figure 7a**) reveals interesting variations.

During photoexcitation, due to changes in the induced dipole moment, the SHG components ratio of anions show a very significant increase, while among the cations, except for Cs which has a slight increase, Sb and S both show significant decreases, especially Sb. By combining the transition matrix diagram with the crystal structure, more fascinating conclusions are discovered.

Given that the transitions with the largest contributions in the structure come from the off-site transitions of O/F−Sb and the on-site transitions of Sb1, the information related to these three components is mainly analyzed. The relevant local structures are divided into two types based on the bonding between O/F and Sb: those that directly bond with Sb and those that do not have strong bonding interactions. Among them, F1, F2, O2, and O4 exhibit strong chemical bond interactions with Sb to form an $SbO_2F_2$ atom group seesaw[14], with Sb1−F1/F2 and Sb1−O2/O4 bond lengths of 1.986/1.957 Å and 2.323/2.165 Å, respectively. In contrast, O1 and O3 exhibit weak chemical bond interactions with Sb, with the closest Sb1···O1 and Sb1···O3 distances being 2.96 and 2.65 Å, respectively.

In the transitions related to Sb and F/O, for the static SHG susceptibility, the relative interatomic contributions of F/O atoms with shorter Sb−F/O bonds are smaller, i.e., $^{proj}\chi_{F2}^{zzz} < {}^{proj}\chi_{F1}^{zzz}$ and $^{proj}\chi_{O4}^{zzz} < {}^{proj}\chi_{O2}^{zzz} < |{}^{proj}\chi_{O3}^{zzz}| < |{}^{proj}\chi_{O1}^{zzz}|$. Among them, the components of O1 and O3, which exhibit weak chemical bonding with Sb1, have opposite signs compared to the macroscopic SHG susceptibility. As can be easily seen from the static SHG susceptibility transition matrix diagram, this is because the dipole moments of O1/O3–Sb1 generate a large contribution $^{proj}\chi_{O1 \to Sb1}^{zzz}$ and $^{proj}\chi_{O3 \to Sb1}^{zzz}$ with opposite sign (**Figure 7b**).

As the intensity of the oscillating light field increases, the macroscopic SHG susceptibility gradually increases, and significant changes occur in the microscopic SHG susceptibility of individual atoms. The proportion of Sb–O transitions components increases more than that of Sb–F transitions. During photoexcitation, O atoms primarily act as charge donors, while F atoms, due to their stronger electronegativity, exhibit significantly weaker charge donor capabilities but slightly stronger charge acceptor capabilities than O (**Figure 6**). Importantly, in case of strong chemical bonds, the SHG components of F/O atoms with shorter Sb−F/O bond lengths becomes relatively larger under photoexcitation, i.e., $^{proj}\chi_{F2}^{zzz} > {}^{proj}\chi_{F1}^{zzz}$ and $^{proj}\chi_{O4}^{zzz} > {}^{proj}\chi_{O2}^{zzz}$ (**Table 1** and **S1**).

In the case of weak chemical bonding, the SHG components from the dynamic O1/O3−Sb1 dipole moments (4.30 × 10$^{-3}$ and 3.95 × 10$^{-3}$ a.u. at 1064 nm) undergo a reversal relative to the zero-frequency limit (−1.57 × 10$^{-3}$ and −6.90 × 10$^{-4}$ a.u.), making these two transitions the strongest contributors to the microscopic SHG susceptibility, in addition to the on-site transitions of Sb1. Furthermore, the SHG components of O1 and O3 atoms decrease from a large gap in the static case (−10.95% vs −2.96%) to a smaller difference (10.65% vs 10.50% at 1064 nm). Unlike in the case of strong bonding, under weak bonding, the SHG components by O atoms with longer Sb···O distances are slightly larger, i.e., $^{proj}\chi_{O1}^{zzz} > {}^{proj}\chi_{O3}^{zzz}$ (**Table 1** and **S1**).

The above analysis shows that even subtle changes in the chemical environment can significantly impact the contributions of the transition dipole moments between atoms to SHG at the zero-frequency limit, and these differences are vastly different from the distribution of dynamic SHG components under actual optical field excitation. These differences can also be visually presented by accumulating all the atomic space SHG distribution functions (**Figure 7c–e**). If one were to solely analyze the contributions of atoms to the static SHG susceptibility in order to explain experimental conclusions, it is possible to obtain unrealistic results, even if it is only used for qualitative analysis.

These pieces of information can be easily extracted through rigorous AST, even including the influence of subtle chemical environments. In other words, the precisely partitioned distribution of wavelength-dependent SHG effects can even accurately respond to minor chemical environmental differences among the same type of chemical bonds in the crystal. This method can directly establish a relationship between the microscopic SHG components of atoms and precise microstructural information in the crystal structure.

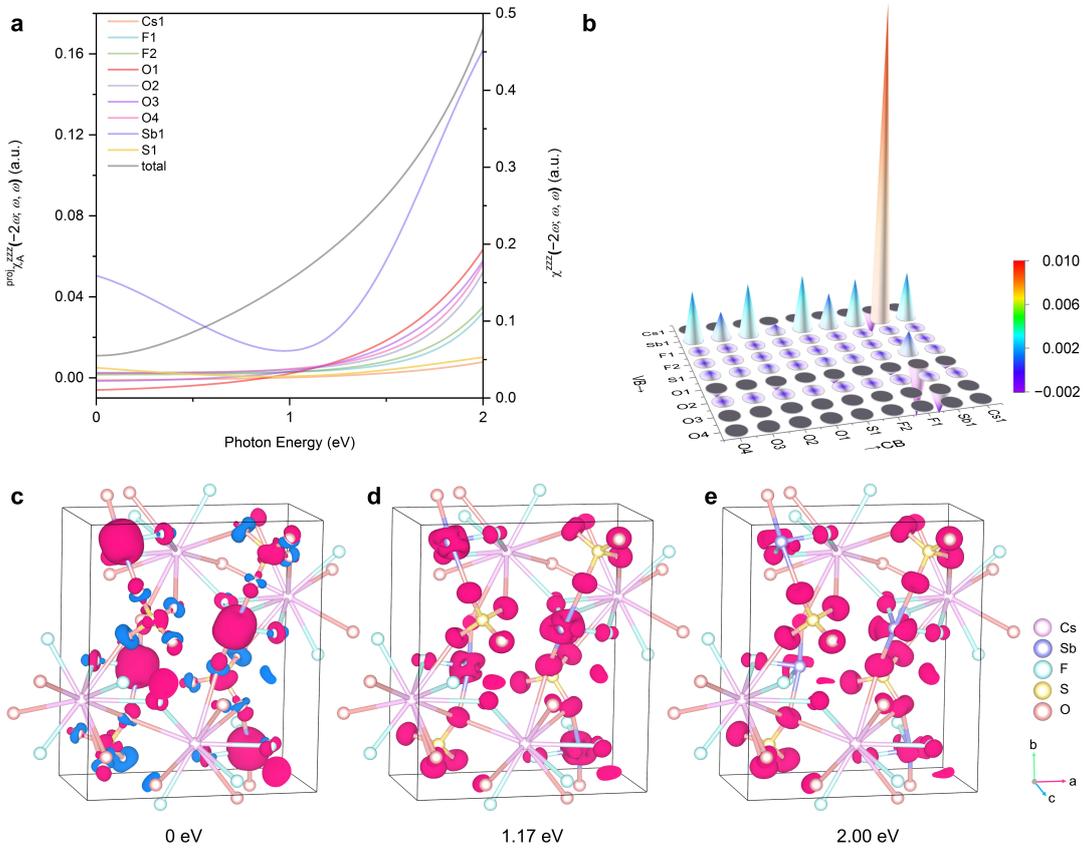

**Figure 7.** The energy-dependent atomic space contribution distribution (**a**) and dipole moment component matrix diagram at the zero-frequency limit (**b**) for $\chi^{zzz}$, obtained through AIM atomic space partitioning method. Sum of all the SHG distribution function in atomic spaces, i.e., $^{proj}\chi^{zzz}(\mathbf{r})$, at different incident photon energies, with isosurface levels of (**c**) 0.00075, (**d**) 0.002, and (**e**) 0.005 Å$^{-3}$, respectively. All units are atomic units.

Many times, visualizing all tensor elements without losing information is expected[24]. More usefully, if all the components of the SHG susceptibility tensor elements in atomic space $\chi_{w_A}^{abc}(-2\omega; \omega, \omega)$ are considered, it becomes possible to examine the contribution of each atom $\chi_A^{(2)}(-2\omega; \omega, \omega)$ to the SHG susceptibility tensor $\chi^{(2)}(-2\omega; \omega, \omega)$ as a whole. In this regard, three-dimensional visualization of the full SHG susceptibility tensor elements is introduced,

$$\chi(\theta,\phi) = \chi : \hat{\boldsymbol{E}}(\theta,\phi)\hat{\boldsymbol{E}}(\theta,\phi). \qquad (8)$$

For CSFS, the $\chi(\theta, \phi)$ at 1064 nm exhibits significant anisotropy. It can be seen that the electric field changing along the polar axis direction of the crystal can obtain the maximum induced dipole moment parallel to the axis, and the same applies to the induced dipole moments contributed by Sb1 and O1 atoms. Differently, in the plane perpendicular to the polarity axis, the induced dipole moments of O1 and unit cell are opposite in direction to the induced dipole moment in the direction of the polarity axis, while Sb1 is the same (**Figure 8**).

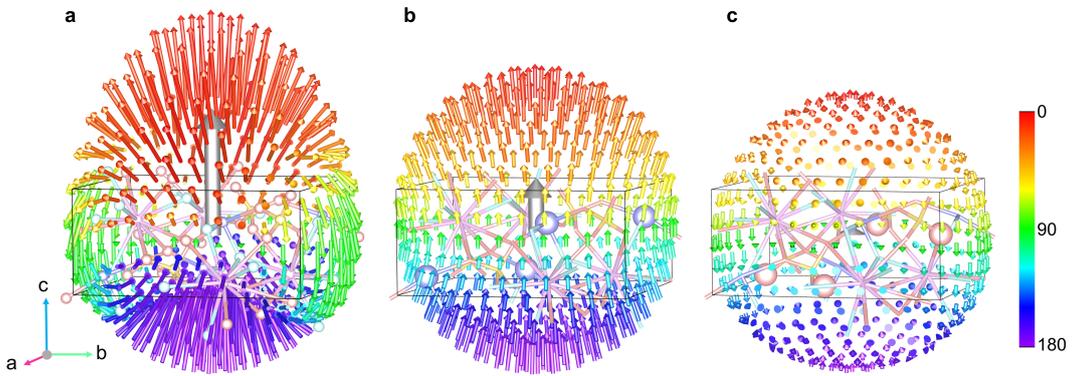

**Figure 8.** $\chi_{CSFS}(\theta, \phi)$, $\chi_{Sb1}(\theta, \phi)$, and $\chi_{O1}(\theta, \phi)$ (arrow color represents the angle between the direction of induced dipole moment and the direction of external electric field change), together with $\chi_{vec}^{CSFS}$, $\chi_{vec}^{Sb1}$, and $\chi_{vec}^{O1}$ (grey arrows), corresponding to (**a**), (**b**), and (**c**), respectively, at an incident wavelength of 1064 nm.

By averaging the coupling effects of electric fields applied in different directions, a SHG susceptibility averaged vector $\chi_{vec}$ can be obtained that ignores anisotropy[25],

$$\chi_{vec} = (\chi_x, \chi_y, \chi_z), \quad \chi_i = \frac{1}{3}\sum_{j=x,y,z}(\chi_{ijj} + \chi_{jij} + \chi_{jji}). \qquad (9)$$

Given that CSFS crystallizes in the polar $C2v$ point group, it follows that $\chi_x = \chi_y = 0$. Therefore, the SHG vector $\chi_{vec}^{CSFS}$ and the components $\chi_{vec}^{A}$ are all parallel to the polar axis. The contribution proportion of the Sb1 component $\chi_{vec}^{Sb1}$ to the total vector $\chi_{vec}^{CSFS}$ ($\chi_{vec}^{Sb1} = 0.46\chi_{vec}^{CSFS}$) is greater than that of Sb1 in a single tensor element $\chi^{zzz}$, which originates from the consistency of coupling responses in different directions (**Figure 8a** and **b**). Conversely, the contribution of the O1 component $\chi_{vec}^{O1}$ to the $\chi_{vec}^{CSFS}$ ($\chi_{vec}^{O1} = 0.075\chi_{vec}^{CSFS}$) is smaller than its contribution proportion within a single tensor element $\chi^{zzz}$. This stems from the inconsistency of coupling responses in different directions, which is more pronounced than that in the total vector (**Figure 8a** and **c**).

**Table 1.** The contribution percentage of each atom to the SHG susceptibility $\chi^{zzz}$ obtained using the AST scheme, at an incident wavelength of 1064 nm.

| Atom | Atomic space partitioning (%) | | | |
|---|---|---|---|---|
| | AIM | Voronoi | Hirshfeld | Hirshfeld-I |
| Cs1 | 1.18 | 0.68 | 3.21 | 1.04 |
| Sb1 | 39.47 | 35.79 | 40.89 | 34.80 |
| S1 | 2.73 | 3.94 | 6.53 | 3.94 |
| F1 | 6.83 | 7.41 | 6.14 | 7.13 |
| F2 | 7.18 | 8.08 | 6.52 | 7.41 |
| O1 | 10.65 | 10.80 | 9.31 | 11.04 |
| O2 | 10.13 | 10.54 | 8.74 | 11.36 |
| O3 | 10.50 | 10.81 | 9.03 | 11.19 |
| O4 | 11.32 | 11.95 | 9.63 | 12.09 |

**Table 2.** The contribution percentage of each atom to the $\Delta\chi^{(1)}$ or $\Delta n^2$ obtained using the AST scheme, at an incident wavelength of 1064 nm.

| Atom | Atomic space partitioning of $\Delta\chi^{(1)}$ or $\Delta n^2$ (%) | | | |
|---|---|---|---|---|
| | AIM | Voronoi | Hirshfeld | Hirshfeld-I |
| Cs1 | −0.43 | −0.24 | 1.47 | −0.02 |
| Sb1 | 37.02 | 32.40 | 38.16 | 31.53 |
| S1 | 2.31 | 3.34 | 6.60 | 3.83 |
| F1 | 0.42 | 0.56 | 0.47 | 1.05 |
| F2 | 1.77 | 2.71 | 1.46 | 2.25 |
| O1 | 16.32 | 16.65 | 14.62 | 16.52 |
| O2 | 15.53 | 16.20 | 13.59 | 16.67 |
| O3 | 15.91 | 16.33 | 13.95 | 16.24 |
| O4 | 11.16 | 12.03 | 9.69 | 11.93 |

The anisotropy of the first-order susceptibilities and the resulting birefringence are also one of the key indicators in the application of NLO crystals. The birefringence of NLO crystals determines whether they can achieve birefringent phase matching. In the application of nonlinear optical crystals, in most cases, the incident and outgoing light are smaller than the bandgap, and the energy loss of the external electric field is very small. Therefore, considering only the real part for the complex form of first-order susceptibility, dielectric function, and refractive index is sufficiently accurate. Similar to the SHG susceptibility, the first-order susceptibility and dielectric function components can be conveniently obtained. Given the relationship between susceptibility and refractive index, an approach is proposed to apply the AST scheme to study the contribution of different atoms to birefringence by analyzing the differences in susceptibility or squares of refractive indices.

The theoretically calculated birefringence is 0.76 at 1064 nm (**Figure S3**), which is a suitable birefringence for phase matching in NLO crystals[7]. The atomic space partitioning results show that Cs and F atoms contribute little to the optical anisotropy of the structure. The sulfate groups contribute significant optical anisotropy, and in addition, Sb also contributes significant optical anisotropy in the same direction as the sulfate (**Table 2**), which enables CSFS to achieve moderate birefringence required for phase matching. Similar to the atomic space SHG distribution function, the polarization anisotropy can also be visualized by sum of all the atomic space $\Delta\chi^{(1)}$ (= $\Delta n^2/4\pi$) distribution functions, i.e., $\Delta\chi^{(1)}$ (**r**), providing a visual representation of the contribution (**Figure 9**).

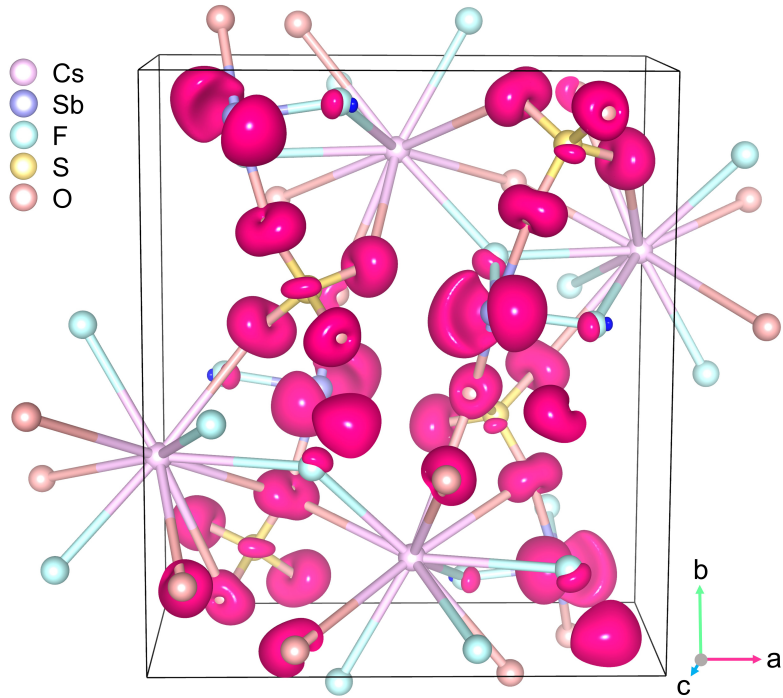

**Figure 9.** $\Delta\chi^{(1)}$(**r**) function isosurface maps at an incident wavelength of 1064 nm (isosurface level = 0.00025 Å$^{-3}$, units in a.u.). The positive and negative isosurfaces are colored in red and blue, respectively.

## CONCLUSION

In summary, the AST scheme, which is used to evaluate the optical components of atoms in crystals, has been implemented. The scheme derives the wavelength-dependent SHG components of various induced dipole moments or individual atoms in nonlinear optical materials without requiring empirical parameters and atomic orbital projections. Furthermore, it possesses well intrinsic stability, meaning that even with different atomic space partitions, the results do not exhibit significant differences. Importantly, the sum of components is naturally normalized. Through the study of CSFS case, the significant difference in the contribution ratio of different atoms to the SHG susceptibility under zero-frequency limit and actual light excitation has been discovered. As a result, it is inaccurate to use static SHG susceptibility components as a substitute for actual SHG susceptibility components under optical excitation for micro-contribution analysis. This inaccuracy extends beyond mere quantitative differences and can even be qualitative. The various analysis techniques extended by the AST scheme reveal the important contribution of Sb to the SHG susceptibilities and anisotropy of the first-order susceptibilities. The implementation of this method may aid in the design of novel nonlinear optical materials with large SHG effects and moderate birefringence.

## AUTHOR INFORMATION

Corresponding Author
Yang Chi –
School of Chemistry and Chemical Engineering, Qufu Normal University, Qufu 273165, P. R. China;
orcid.org/0000-0002-7021-3891;
Email: yang.chi@hotmail.com

Notes
The authors declare no competing financial interest.

## ACKNOWLEDGMENTS

The authors acknowledge financial support by the National Science Foundation of China (22305142), and the Natural Science Foundation of Shandong Province (ZR2021QB017).

# *Supporting Information*

**Evaluation of Microscopic Origins in Nonlinear Optical Crystals: Based on Rigorous Atomic Space Tessellating**

*Yang Chi\**

*School of Chemistry and Chemical Engineering, Qufu Normal University, Qufu 273165, P. R. China*

*Corresponding author: yang.chi@hotmail.com*

# I. Susceptibilities

The following expressions with scissor correction derived within the length-gauge formalism are used to calculate the frequency-dependent SHG susceptibility in an insulator or clean semiconductor at zero temperature[1],

$$\chi^{abc}(-2\omega;\omega,\omega) = \chi_e^{abc}(-2\omega,\omega,\omega) + \chi_i^{abc}(-2\omega,\omega,\omega), \tag{S1}$$

$$\chi_e^{abc}(-2\omega;\omega,\omega) = \frac{e^3}{\hbar^2 \Omega} \sum_{nml,\mathbf{k}} \frac{r_{nm}^a \{r_{ml}^b r_{ln}^c\}}{(\omega_{ln} - \omega_{ml})} \times \left[ \frac{2f_{nm}}{\omega_{mn} - 2\omega} + \frac{f_{ln}}{\omega_{ln} - \omega} + \frac{f_{ml}}{\omega_{ml} - \omega} \right], \tag{S2}$$

$$\chi_i^{abc}(-2\omega;\omega,\omega) = \frac{i}{2}\frac{e^3}{\hbar^2 \Omega} \sum_{nm,\mathbf{k}} f_{nm} \left[ \frac{2}{\omega_{mn}(\omega_{mn} - 2\omega)} r_{nm}^a (r_{nm;c}^b + r_{nm;b}^c) + \frac{1}{\omega_{mn}(\omega_{mn} - \omega)} (r_{nm;c}^a r_{mn}^b + r_{nm;b}^a r_{mn}^c) \right.$$
$$\left. + \frac{1}{\omega_{mn}^2} \left( \frac{1}{\omega_{mn} - \omega} - \frac{4}{\omega_{mn} - 2\omega} \right) r_{nm}^a (r_{mn}^b \Delta_{mn}^c + r_{mn}^c \Delta_{mn}^b) - \frac{1}{2\omega_{mn}(\omega_{mn} - \omega)} (r_{nm;a}^b r_{mn}^c + r_{nm;a}^c r_{mn}^b) \right]. \tag{S3}$$

The calculation of the first-order susceptibility tensor adopts a well-known form,

$$\chi_{ab}^{(1)}(-\omega;\omega) = \frac{e^2}{\hbar\Omega} \sum_{n,m,\mathbf{k}} \frac{r_{nm}^a r_{mn}^b f_{mn}}{\omega_{nm} - \omega}. \tag{S4}$$

The real part $n$ of the complex refractive index is calculated using the following formula,

$$n(\omega) = \frac{1}{\sqrt{2}} \left[ (\varepsilon_1(\omega)^2 + \varepsilon_2(\omega)^2)^{1/2} + \varepsilon_1(\omega) \right]^{1/2}. \tag{S5}$$

# II. Space partition methods in the AST scheme

## 1. Discrete atomic space analysis

There are two most representative methods here, Voronoi[2] (cell-like) and "the quantum theory of atoms in molecules"[3] (AIM) partition, which were originally proposed to analyze the charge density of molecules. Both methods are partition space discretely, so any point in Bloch orbital can be attributed to only one atom,

$$\begin{cases} w_A(\mathbf{r}) = 1, & \mathbf{r} \in \Omega_A \\ w_A(\mathbf{r}) = 0, & \mathbf{r} \notin \Omega_A \end{cases},$$

where $\Omega_A$ is atomic space of atom A.

Therefore, the sum of all atomic spaces in the unit cell is strictly equal to the unit cell volume (for AIM partitioning, in the absence of pseudoatoms),

$$\sum_i \Omega_A^i = \Omega.$$

The AIM partition adopts a physically meaningful approach, dividing the entire three-dimensional space into atomic basins through the zero-flux surface of electron density, with no electron density gradient lines crossing the interface. These independent spaces correspond to the atomic spaces defined by AIM theory. In contrast, Voronoi atomic spaces rely solely on the crystal structure, forming closed polyhedra through perpendicular bisectors with neighboring atoms. A characteristic of Voronoi atomic spaces is that any point within a Voronoi atomic space is closer to the atom in that space than to any other atom. In other words, any point in three-dimensional space is assigned to the Voronoi atomic space of the nearest atom.

## 2. Fuzzy atomic space analysis

The Hirshfeld partition[4] is one of the widely used fuzzy space methods, and subsequently, many other fuzzy space partition methods such as Becke[5] and Hirshfeld-I[6] have been proposed. They continuously partition the three-dimensional space, and the atomic spaces divided by these methods overlap with each other. From a three-dimensional perspective, all atoms occupy exactly the same space, which is the entire molecular or crystal space. Any point in the entire three-dimensional space is attributed to any atom, but with different weights for different atoms. All atoms and any point satisfy the following two conditions,

$$\begin{cases} 0 \leq w_A(\mathbf{r}) \leq 1, \quad \forall A \\ \sum_i w_A^i(\mathbf{r}) = 1 \end{cases}.$$

In order to ensure the determinacy of atomic space partitioning and the absence of empirical parameters, Hirshfeld and Hirshfeld-I methods were chosen here.

Hirshfeld atomic space is defined as[4],

$$w_A^{\text{Hirshfeld}}(\mathbf{r}) = \frac{\rho_A(\mathbf{r})}{\rho^{\text{pro}}(\mathbf{r})},$$

where $\rho^{\text{pro}}$ is procrystal electron density, $\rho_A$ denotes spherically averaged atomic electron density in free state.

Given that the Hirshfeld atomic space does not respond to the actual chemical environment around the atom, subsequent methods have been proposed to iteratively adjust the atomic spaces in response to the surrounding environment. These methods are considered more physically meaningful than the Hirshfeld ones. Among the atomic iteration methods, a representative one is the Hirshfeld-I method, which is an important extension of the Hirshfeld method. The atomic charges obtained through this method are more in line with chemical intuition than those obtained by the Hirshfeld method.

Hirshfeld-I atomic space is defined as[6],

$$w_A^{\text{Hirshfeld-I},n}(\mathbf{r}) = \frac{\rho_A^{n-1}(\mathbf{r})}{\rho_{pro}^{n-1}(\mathbf{r})},$$

$$\rho_A(\mathbf{r}) = \rho_A^{\text{int}(N_A)}(\mathbf{r}) + \left[\rho_A^{\text{int}(N_A)+1}(\mathbf{r}) - \rho_A^{\text{int}(N_A)}(\mathbf{r})\right]\left[N_A - \text{int}(N_A)\right],$$

where, the notations int($N_A$) to express the integer part of the atomic electronic population was used.

## III. Numerical computation

Structural optimization and electronic structure calculations of CSFS were carried out within the framework of density functional theory (DFT) by using the *CASTEP* package[7]. The generalized gradient approximation (GGA) within the Perdew−Burke−Ernzerhof (PBE)-type exchange-correlation potentials were used throughout this work[8]. The employed OTFG norm-conserving pseudopotentials of Cs, Sb, F, S, and O treat 5s 5p 6s, 4d 5s 5p, 2s 2p, 3s 3p, and 2s 2p as the valence states, respectively. The plane-wave cutoff energy of 1000 eV and the threshold of $5 \times 10^{-7}$ eV/atom were set for the self-consistent-field convergence of the total electronic energy. The atomic positions were allowed to relax to minimize the internal forces. An excellent convergence of the energy differences ($5.0 \times 10^{-6}$ eV/atom), maximum force (0.01 eV/Å), and maximum displacement ($5.0 \times 10^{-4}$ Å) was implemented in the atomic position optimization. A $5 \times 4 \times 9$ Monkhorst−Pack *k*-point grid in the Brillouin Zone of the primitive cell are chosen and more than 545 empty bands were involved in the calculations to ensure the convergence of SHG coefficients.

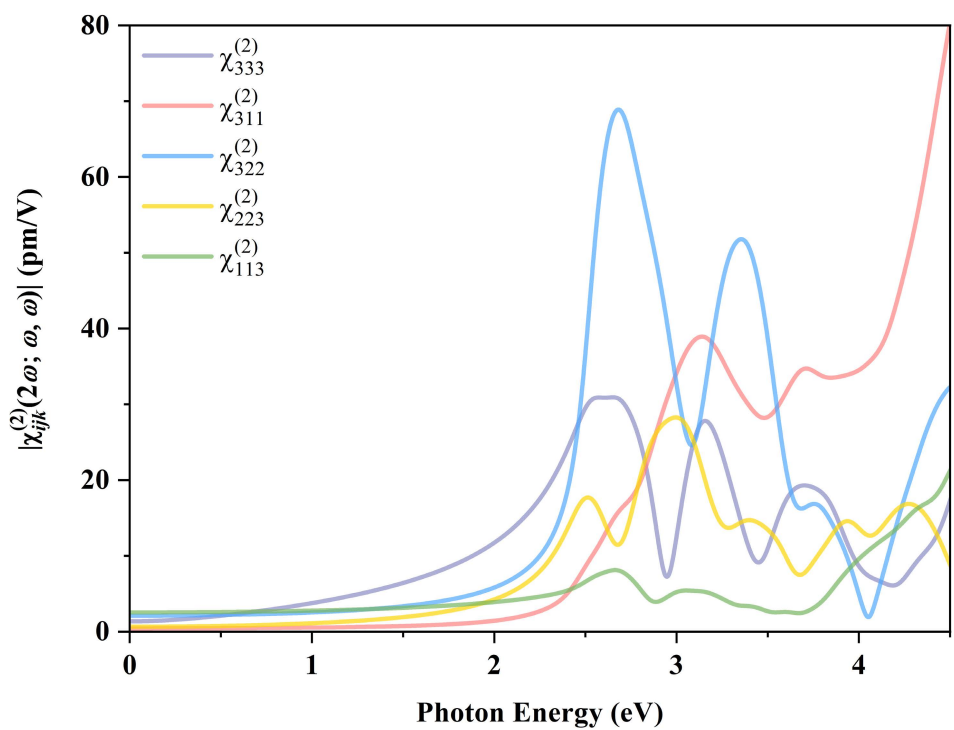

**Figure S1.** Frequency dependency of $|\chi^{(2)}_{ijk}(-2\omega; \omega, \omega)|$ for CSFS.

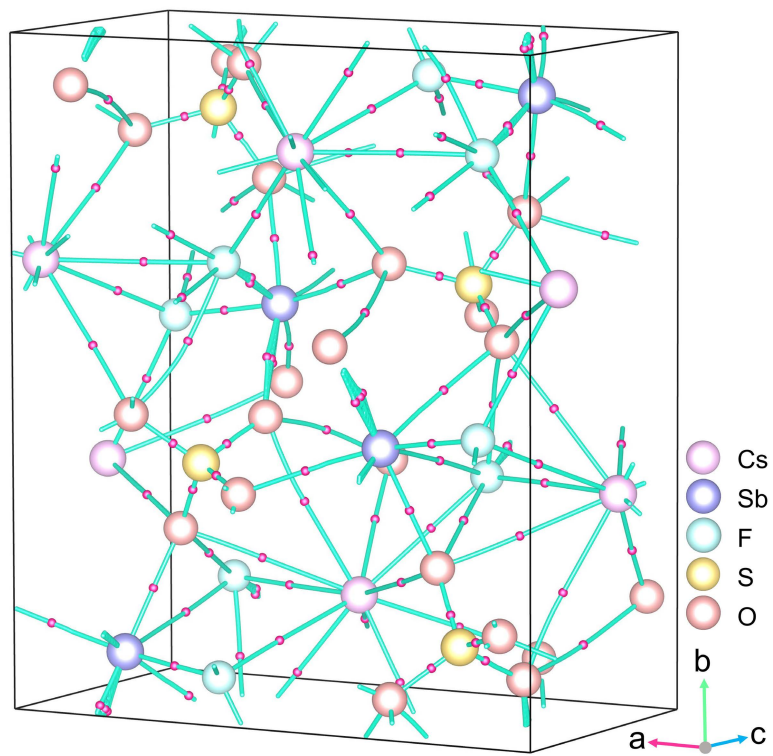

**Figure S2.** Bond path (green line) and critical point (rose red ball) in CSFS.

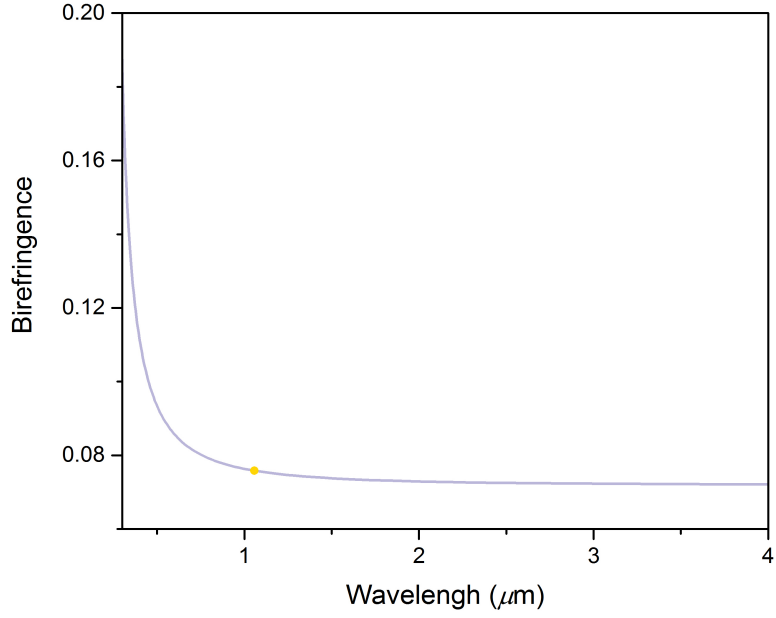

**Figure S3.** Calculated birefringence of CSFS. The value at 1064 nm is marked with a yellow dot.

**Table S1.** The contribution percentage of each atom to the SHG susceptibility $\chi^{zzz}$ ($-2\omega; \omega, \omega$) obtained using the AIM partition in the AST scheme, at incident light energies of 0 and 2.0 eV, respectively.

| Atom | $^{proj}\chi^{zzz}_{AIM-A}(-2\omega; \omega, \omega)/\chi^{zzz}(-2\omega; \omega, \omega)$ (%) | |
|---|---|---|
| | 0 eV | 2.00 eV |
| Cs1 | −2.00 | 1.64 |
| Sb1 | 91.54 | 33.36 |
| S1 | 9.00 | 2.10 |
| F1 | 4.09 | 6.90 |
| F2 | 2.81 | 7.46 |
| O1 | −10.95 | 13.30 |
| O2 | 4.61 | 11.10 |
| O3 | −2.96 | 12.18 |
| O4 | 3.87 | 11.97 |